\documentclass[journal]{IEEEtran}

\usepackage{blindtext}
\usepackage{graphicx}
\usepackage{amsmath}
\usepackage{mathpazo}	
\usepackage{tikz}
\usepackage{color}
\usepackage{cite}
\usepackage{subfig}
\usepackage{lipsum}

\ifCLASSINFOpdf
\else
\fi
\begin{document}
%

\title{Perturbation-based Frequency Domain Linear and Nonlinear Noise Estimation}

%
%
%

\author{{
    F.J. Vaquero-Caballero,     
	D.J. Ives,
 	S.J. Savory
}
%
%


\thanks{FJVC (e-mail: fjv24@cam.ac.uk), DJI, and SJS are the Electrical Engineering Division, Department of Engineering, University of Cambridge, Cambridge CB3 0FA, UK.}	
\thanks{\textcolor{red}{Manuscript received February 26, 2021}}}

%
%

\markboth{Journal of Lightwave Technology}%
{Shell \MakeLowercase{\textit{et al.}}: Bare Demo of IEEEtran.cls for IEEE Journals}
%



\maketitle

\begin{abstract}
In this paper, a new method for the separation of noise categories based on Four-Wave Mixing is presented. 

The theoretical analysis is grounded in the Gaussian Noise model and verified by split step simulations. The noise categories react differently to the introduced perturbations, by performing a set of perturbations the behaviour of the different categories can be separated by means of a least-square fitting. Given ASE is independent of the induced perturbations, it is possible to separate noise contributions. The analysis includes constant and variable power perturbations. 

The estimation of the noise categories is discussed from two points of view: NSR evolution post-DSP processing, and over the power spectral density in a notched region. The NSR estimation can only be performed at reception, whereas the power spectral density approach can be performed along the optical link if a high resolution Optical Spectrum Analyzer is available.

Additionally, we perform a simple experimental verification considering of two WaveLogic 3 transceivers for the NSR, successfully estimating the noise contributions. 
\end{abstract}

\begin{IEEEkeywords}
IEEE, IEEEtran, journal, \LaTeX, paper, template.
\end{IEEEkeywords}

%
\IEEEpeerreviewmaketitle

\section{Introduction}
Motivated by increasing throughput demand, network operators seek to capitalize margin stranded by guard-bands that are put in place to buffer uncertain knowledge of link state. To improve link state knowledge an area of interest is the measurement and partitioning of channel noises.

Some proposed techniques of noise partitioning are offline: \cite{Ives2019} abstracting the network elements to an noise-to-signal ratio (NSR) terms, allowing fast computation of the expected performance. Similarly,\cite{Chin2017} proposed the application of probabilistic design to combine noise from independent origins and reduce guard band margins needed for a chosen risk of failure.

In service noise measurements can be performed on extracted  information from the transmit signal \cite{CaballeroVaquero2018,Dong2012,Caballero18,Vaquero-Caballero2019,Liu2020,Shiner2020,Gariepy2018,Lonardi2020,Tanimura2016,Khan2017,Cho2019,Cho2020}. We have proposed a time domain method using normal and tangential noise components of a constellation to measure a non-linear phase noise distinct from a contributing circular symmetric noise, to infer linear and nonlinear noise (NLN),\cite{CaballeroVaquero2018}. Ref. \cite{Dong2012} has proposed to exploit the temporal correlation of certain nonlinearities to determine a correction factor in the Error-Vector Magnitude (EVM) Signal-to-Noise Ratio (SNR) and disambiguate an OSNR estimate. Other investigations have been inspired by this approach: in \cite{Caballero18}, we proposed to measure the second order statistics of the nonlinearities and use a small neural network to estimate the ASE and the NLN contributions. Given long temporal correlation, Principal Component Analysis (PCA) reduced the dimensionality of the metrics. In \cite{Vaquero-Caballero2019} we explored DSP dependencies of the time domain metrics. Similar approaches based on temporal correlations, are currently investigated as a way to segregate intra- and inter-channel non-linear noise, \cite{Liu2020}. Finally, Ciena WaveLogic transceivers incorporate a similar time-domain implementation to provide real time estimates of OSNR \cite{Shiner2020}. 

In frequency domain, ref. \cite{Gariepy2018} obtain a noise partitioning based a comparison of the received spectra w.r.t. a reference spectrum, using a commercial OSA. The separation of noises between ASE and NLN is made possible by the spectral flatness of ASE, when compared with the NLN. Similarly \cite{Lonardi2020}, proposed a pseudo-analytical method to estimate the noise-to-signal spectrum based on an incoherent GN model. The method of \cite{Gariepy2018} operates predominantly on the spectral transitions at high and low frequency edges and, as such, is vulnerable to the action of optical filtering present in the optical line.

Deep Neural networks have also being explored for the segregation of linear and nonlinear noise: \cite{Tanimura2016,Khan2017,Wang2018a,Cho2019,Cho2020}.

To date linear and nonlinear noise estimation have been mainly implemented using machine learning, based on phenomenological relationships, and service- optimized transmission spectrum. In contrast here we, introduce advantageous perturbations of the transmitted spectrum and analyze their effect on the received spectrum to partition noises. 
\section{Mathematical Background}
This section introduces a mathematical framework of the proposed method: the GN model, the NLN classification terminology, and an averaged spectral term.

Since there is not a canonical form for the NLN classification terminology, our nomenclature is inspired by \cite{Lang1997} where a similar approach was used to study Volterra kernel generated NLN. Our terminology also includes polymorphism: functions with different meaning depending on the number of input arguments, e.g.: $X(A)$, $X(A,B)$.

\subsection{The GN model}
The Gaussian Noise (GN) model, is a well-recognized and verified model to describe nonlinearities assuming independence between the frequency components of the signal.

Despite discrepancies between the GN and the experimental results for low transmission distances, low accumulated dispersion, and low cardinality modulation formats; for the typical conditions of optical transmission, it provides an adequate representation for our purpose of nonlinearity as a noise power. For a transmitted optical power spectral density, $S_{Tx}(f)$, the nonlinear noise power spectral density, $S_{NLN}(f)$, can be approximated as:
\begin{equation}
\vspace{-0.5em}
\begin{aligned}
\begin{split}
S_{NLN}(f)= 
	\int_{-\infty}^{\infty} \!\!\!  \, \int_{-\infty}^{\infty} S_{Tx}(f_1)  \, S_{Tx}(f_2) \, S_{Tx}(f_1+f_2-f) \\ FWMW(f_1,f_2,f,\Theta) \, df_1 \, df_2, 
\end{split}
\end{aligned}
\label{eq:NLN_GN_Int}
\end{equation}
where $FWMW(f_1,f_2,f,\Theta)$ is a Four Wave Mixing Weight that depends on link details such as topology, power profile, and dispersion map, among others.

Equation \ref{eq:NLN_GN_Int} is obtained from \cite{Poggiolini2014}, with $\Theta$ representing all link parameter dependencies. For simplicity and without loss of generality we confine our analysis to isotropic links (identical spans and launch power conditions).

Our analysis is focused on frequency domain, but it is also possible to derive similar conclusions in time domain \cite{Dar2013}.
\subsection{Nonlinear Noise Frequency Classification}

In four wave mixing (FWM) 3 frequency components: $f_1$, $f_2$, and $f_1+f_2-f$; interact in a medium to produce a component at a fourth frequency: $f$.

If a spectrum is divided into two regions, $F_A$ and $F_B$, it is possible to categorize the nonlinear spectrum according to constituent generating frequency field regions by conditioning the double integral of Equation \ref{eq:NLN_GN_Int} to the defined frequency ranges. One obtains 8 categories following a permutation with repetition rule:
\begin{equation}
\begin{split}
S_{NLN}(f) =
\\
S_{NLN}'(A,A,A,f)	+	S_{NLN}'(A,A,B,f) \\
+S_{NLN}'(A,B,A,f) 	+	S_{NLN}'(A,B,B,f)\\
+S_{NLN}'(B,A,A,f) 	+	S_{NLN}'(B,A,B,f)\\
+S_{NLN}'(B,B,A,f) 	+	S_{NLN}'(B,B,B,f),\\
\end{split}
\label{eq:NLNfreqFragPR}
\end{equation}
where the first 3 arguments in $NLN'$ are frequency ranges of the transmitted spectrum. I.e.: $S_{NLN}'(A,B,C,f)$ would correspond to the integral of Equation \ref{eq:NLN_GN_Int}, conditioned on $f_1 \in F_A$, $f_2 \in F_B$, and $f_1+f_2-f \in F_C$.

Given we are concerned about the power multiplicity of the frequency categories, a multiset classification instead of permutation  with repetition rule classification is required, thus Equation \ref{eq:NLNfreqFragPR} becomes:
\begin{equation}
\begin{split}
S_{NLN}(f) =	
\\
S_{NLN}(A,A,A,f) + S_{NLN}(B,A,A,f)
\\
+ S_{NLN}(B,B,A,f)	+ S_{NLN}(B,B,B,f),
\end{split}
\label{eq:NLNfreqFragMS}
\end{equation}
As an example, the terms on the LHS\footnote{The definitions of the total nonlinear spectrum: $|NLN(f)|^2$, and the NLN multiset classification, e.g.: $|NLN(A,A,A,f)|^2$, share the same terminology, and are distinguishable by the number of arguments used.} have the following meanings:
\begin{equation}
\begin{split}
S_{NLN}(A,A,A,f)    =  S_{NLN}'(A,A,A,f),	\\
S_{NLN}(B,A,A,f)    =  S_{NLN}'(B,A,A,f) +	\\
 S_{NLN}'(A,B,A,f)+ S_{NLN}'(A,A,B,f),
\end{split}
\end{equation}

\subsection{Average Power Spectral Density and NSR notation}
In this paper, the power of spectral regions will be changed to infer the NLN and the ASE contributions. For compactness, we define the Average Power Spectral Density\footnote{To easy the reading of the notation, $F_{rg}$ is also abbreviated to $rg$ for use in superscripts or subscripts; and subscripts in spec are written without subscripts, e.g. $\overline{P}_{SPEC_1}^{(F_{rg})}= \overline{P}_{SPEC,1}^{(rg)}$.} (APSD) over a frequency range $F_{rg}$ of a spectrum $|SPEC(f)|^2$ as:
\begin{equation}
\overline{S}_{TX}^{(rg)} = \frac{\int_{<F_{rg}>} S_{Tx}(f) \, df}{\int_{<F_{rg}>} \,df},
\end{equation}
For example, the received spectrum, $|RX(f)|^2$, can be written as the addition of the transmitted power spectrum $S_{Tx}(f)$, Amplified Spontaneous Emission Noise power spectrum $S_{ASE}(f)$, and FWM power spectrum $S_{NLN}(f)$:
\begin{equation}
\begin{split}
S_{Rx}(f) = S_{Tx}(f) +	S_{NLN}(f)+ S_{ASE}(f)
\label{eq:NoiseAddition}
\end{split}
\end{equation}
{Where it is assumed that noises of these power spectra are uncorrelated, that assumption will be maintained during the totality of this paper}. The APSD addition of Equation \ref{eq:NoiseAddition} for $F_{rg}$ is:
\begin{equation}
\begin{split}
\overline{S}_{RX}^{(rg)} = \overline{S}_{TX}^{(rg)} + \overline{S}_{NLN}^{(rg)} + \overline{S}_{ASE}^{(rg)},
\end{split}
\label{eq:NoiseAdditionAPSD}
\end{equation}
Where it is assumed that noises of these power spectra are uncorrelated. For the case of nonlinear noise elements, the nomenclature used to refer to their APSD is:
\begin{equation}
\overline{S}_{NLN}^{(rg)}(A,A,A)	=	\frac{\int_{<F_{rg}>} S_{NLN}(A,A,A,f)  \, df }{\int_{<F_{rg}>} \,df},
\end{equation}

Similarly a NSR notation identifies the ratio of the NLN category power to the signal power in the bandwidth of the signal $F_{signal}$:
\begin{equation}
NSR_{NLN}^{(signal)}(A,A,A)	=	\frac{\overline{S}_{NLN}^{(signal)}(A,A,A)} { \overline{S}_{Tx}^{(signal)}},
\end{equation}
without lose of generality, previous noise terms may have been filtered by a receiver match filter.
\begin{figure}
\vspace{-4mm}
\centering
\begin{tikzpicture}
\draw	[thick,draw=black] 	(-1,2)		rectangle	(-0.5,0);
\draw	[thick,draw=black] 	(0.5,1)		rectangle	(1,0);
\draw	[thick,dashed,draw=red,] 	(-1,1.5) 	rectangle	(-0.5,0);
\draw	[thick,dashed,draw=red] 	(0.5,1.5)	rectangle	(1,0);

\node	[red] at (-1.5,1.7)	{$\Delta_k(A)$};
\node	[red] at ( 1.5,1) 	{$\Delta_k(B)$};

\draw 	[->,thick]		( -3,0)		--	(3,0)		node[right=-0.1] {$f$};		
\draw 	[->,thick]		(0,0)		-- 	(0,2.25)	node[above] {$S_{Tx,k}(f)$};	
\draw	[<->,thick] 	(-1,-0.5)	--	node[above] {$F_{A}$} (-0.5,-0.5);
\draw 	[thick,dotted]	(-1,-1) 	--	(-1,0);
\draw 	[thick,dotted]	(-0.5,-0.5)	--	(-0.5,0);
\draw	[<->,thick] 	(0.5,-0.5)	--	node[above] {$F_{B}$} (1,-0.5);
\draw 	[thick,dotted]	(0.5,-0.5) 	--	(0.5,0);
\draw 	[thick,dotted]	(1,-1)		--	(1,0);
\draw	[<->,thick] 	(-0.5,-0.5)	--	node[above] {$F_{N}$} (0.5,-0.5);
\draw	[<->,thick] 	(-1,-1)		--	node[above] {$F_{BOI}$} (1,-1);
\node	[red] at (1.1,2) {$S_{Tx,ref}(f)$};

\draw	[thick]			( 1.35,	0)	--	(1.35,1.5);	
\draw	[thick]			( 1.35,1.5)	--	(2.75,1.5);	
\draw	[dotted,thick]	( 2.75,1.5)	--	(3,1.5);

\draw	[thick]			(-1.35,	0)	--	(-1.35,1.5);
\draw	[thick]			(-1.35,1.5)	--	(-2.75,1.5);
\draw	[dotted,thick]	(-2.75,1.5)	--	(-3,1.5);

\draw	[->,thick] 		(-2.75,-1)	--	node[above] {$F_{OB}$} 	(-1,-1);
\draw	[<-,thick] 		(1,-1)		--	node[above] {$F_{OB}$}	(2.75,-1);
\draw	[dotted,thick]	(-2.75,-1)	--	(-3,-1);
\draw	[dotted,thick]	(2.75,-1)	--	(3,-1);
\end{tikzpicture}
\caption{Reference spectrum and perturbed spectrum with ouf-of-band signal.}
\label{fig:PertSimplExampleTotal}
\end{figure}
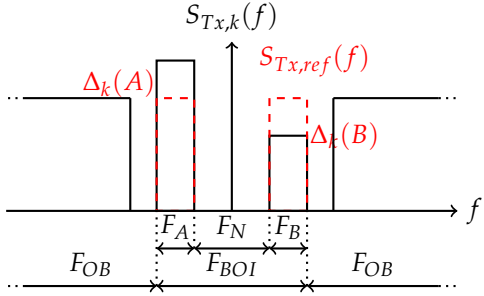
\section{Methodology}
The proposed noise segregation technique is suitable for the transmission of an empty sepctra region with no energy associated and for the NSR calculation of individually modulated subcarriers.

Figure \ref{fig:PertSimplExampleTotal} shows an example of noise partitioning power spectra. The vertical axis is linear in power spectral density. Shown are 2 transmitter power spectra, one reference and one (of many) perturbed from the reference: $S_{Tx,ref}(f)$, and $S_{Tx,k}(f)$, respectively. These simplified spectra share three spectral regions: $F_A$, $F_B$, and a notch region in between with zero power $F_N$. A Bandwidth of Interest (BOI) is the union of $F_A$, $F_B$, and $F_N$, which corresponds to the region of the spectrum we can control. $F_{OB}$ is the out-of-band region of the spectrum.

The perturbed spectrum $S_{Tx,k}(f)$ can be obtained from the reference spectrum, $S_{Tx,ref}(f)$, by multiplicative terms (gain/loss) $\Delta_k(A)$ and $\Delta_k(B)$, where $k$ indexes several realizations of the perturbed spectrum. $\Delta_k(A)$ and $\Delta_k(B)$ operate on frequency ranges $F_A$ and $F_B$, respectively:
\begin{equation}
S_{Tx,k}(f)	=	
 \begin{cases}
S_{Tx,ref}(f) \Delta_k(A), \quad f \in F_A,\\
S_{Tx,ref}(f) \Delta_k(B), \quad f \in F_B,
 \end{cases}
\label{eq:DeltaDef}
\end{equation}
In the following subsections, we discuss methods for the separatation the noise components based on an NSR of a data carried signal or the eploitation of zero energy regions of the spectrum based on APSD approach. We also introduce a constant power noise segregation approach for systems where power variations can disrupt the normal mode of operation.
\subsection{NSR-based Segregation}
if $F_A$ is a data carrying signal whose NSR can be extracted from the receiver DSP, the NSR depencency with $\Delta_1(A)$ for the [A,A,A] category is:
\begin{equation}
NSR^{(A)}_1(A,A,A) = \Delta_1^2(A) NSR^{(A)}_{NLNref}(A,A,A),
\end{equation}
Considering all NSR terms for a set of $p$ perturbations, the received NSRs, $\overrightarrow{NSR}_{Rx}^{(A)}$, are a function of the reference NSRs, $ \overline{\overline{\Delta}}_{NSR}$, and a perturbation delta matrix, $ \overrightarrow{NSR}_{REF}^{(A)}$:
\begin{equation}
\overrightarrow{NSR}_{Rx}^{(A)} = \overline{\overline{\Delta}}^{(A)}_{NSR} \, \overrightarrow{NSR}_{REF}^{(A)},
\end{equation}
where $\overrightarrow{NSR}^{(A)}$ is the received NSR vector for the $p$ realizations of the perturbations:
\begin{equation}
\overrightarrow{NSR}^{(A)} 	=	[NSR_1^{(A)},NSR_2^{(A)},\cdots,NSR_p^{(A)}]^T,
\end{equation}

 $\overrightarrow{NSR}_{REF}^{(A)}$ indexes the NSR of the NLN categories, $NSR_{NLNref}$ , the transceiver NSR, $NSR_{TRX}$, and the ASE NSR, $NSR_{ASE}$:
\begin{equation}
\begin{split}
 \overrightarrow{NSR}_{REF}^{(A)}=	\\
[NSR_{NLNref}^{(A)}(A,A,A),NSR_{NLNref}^{(A)}(B,A,A), \cdots
\\
 NSR_{NLNref}^{(A)}(B,B,A),NSR_{NLNref}^{(A)}(B,B,B),\cdots \\ NSR_{TRX} ,NSR_{ASE}]^T,
\end{split}
\end{equation}

And each row of the $\overline{\overline{\Delta}}_{NSR}^{(A)}$ indexes the influence of $\Delta(A)$ and $\Delta(B)$ over the NSR categories:
\begin{equation}
\small
\begin{split}
\overline{\overline{\Delta}}_{NSR}^{(A)} = 
\\
\begin{bmatrix}
\Delta_1^2(A)	&	\Delta_1(A)	\Delta_1(B) & \Delta_1^2(B)	& \Delta_1^3(B)/\Delta_1(A)	&	1 		&	1/\Delta_1(A)	\\
\Delta_2^2(A)	&	\Delta_2(A)	\Delta_1(B) & \Delta_2^2(B)	& \Delta_2^3(B)/\Delta_1(A)	&	1 		&	1/\Delta_2(A)	\\
\vdots 			&	\vdots 					& 	\vdots 		&	\vdots					&	\vdots	&	\vdots	\\
\Delta_p^2(A)	&	\Delta_p(A)	\Delta_p(B) & \Delta_p^2(B)	& \Delta_p^3(B)/\Delta_p(A)	&	1 		&	1/\Delta_p(A)	\\
\end{bmatrix},
\end{split}
\label{eq:DeltaMAT}
\end{equation}

Without lose of generality, we only present the NLN categories generated by $F_A$ and $F_B$, in $F_A$. This defintions are suitable to any arbitrary perturbation topology, including $F_{OB}$ contributions. Contributions in $F_{OB}$ are assumed to be constant $\Delta(OB)=1$.

Thus, the least-square estimation is:
\begin{equation}
\overrightarrow{NSR}_{REF}^{(A)} = \Big( (\overline{\overline{\Delta}}_{NSR}^{(A)})^{T} \overline{\overline{\Delta}}_{NSR}^{(A)} \Big)^{-1} (\overline{\overline{\Delta}}_{NSR}^{(A)})^{T} \overrightarrow{NSR}^{(A)},
\label{eq:FittingCaseSubCarrier}
\end{equation}
The data carrying signal in $F_B$ accordingly.

\subsection{APSD-based Segregation}
Similarly to the NSR case, the dependency with $\Delta_1(A)$ of the [A,A,A]  category in $F_N$:
\begin{equation}
\overline{S}_{NLN}^{(rg)}(A,A,A) = \Delta_1^3(A) \overline{S}_{NLNref}^{(rg)}(A,A,A),
\end{equation}

For the case of the region $F_N$, the received APSD , $\overrightarrow{P}_{RX}^{(N)}$ is a function of the reference APSD vector, $\overrightarrow{P}_{REF}^{(N)}$, and the perturbation delta matrix, $ \overline{\overline{\Delta}}_{APSD}$:
\begin{equation}
\overrightarrow{S}_{RX}^{(N)} = \overline{\overline{\Delta}}_{APSD} \, \overrightarrow{S}_{REF}^{(N)}
\end{equation}

Assuming transceiver contribution is neglible, $\overrightarrow{P}_{RX}^{(N)}$ and $\overrightarrow{P}_{REF}^{(N)}$ are vectors that indexes received and reference APSDs:
\begin{equation}
\begin{split}
\overrightarrow{S}_{RX}^{(N)}=[\overrightarrow{S}_{RX,1}^{(N)},\overrightarrow{S}_{RX,2}^{(N)},\cdots, \overrightarrow{S}_{RX,p}^{(N)}]^T,
\\
\overrightarrow{S}_{REF}^{(N)}=[\overline{S}_{NLNref}^{(N)}(A,A,A), \overline{S}_{NLNref}^{(N)}(B,A,A), \cdots \\ 
\overline{S}_{NLNref}^{(N)}(B,B,A), \overline{S}_{NLNref}^{(N)}(B,B,B),\overline{S}_{ASE}^{(N)}]^T,
\end{split}
\end{equation}

and the $k$ row of $\overline{\overline{\Delta}}_{APSD}$ is:
\begin{equation}
[\Delta_k^3(A),\Delta_k^2(A)\Delta_k(B),\Delta_k(A)\Delta_k^2(B),\Delta_k^3(B)],
\end{equation}

Similarly to the NSR case, the least-square estimation is:
\begin{equation}
\overrightarrow{S}_{REF}^{(N)} = ((\overline{\overline{\Delta}}_{APSD})^{T} \overline{\overline{\Delta}}_{APSD})^{-1} \overline{\overline{\Delta}}_{APSD}^{T} \overrightarrow{S}_{RX}^{(N)},
\label{eq:FittingCase}
\end{equation}

\subsection{Constant Power Perturbation}
Changing $|TX|^2$ power can change the state of the optical link, given that amplifiers and other optical components can be affected. Thus, it is possible to define a constant power perturbation as:
\begin{equation}
\Delta(B) 	= \frac{1-K_{A} \Delta(A)}{K_B},
\label{eq:ConstPowerContr}
\end{equation}	
where $K_A$ and $K_B$ are the relative power contributions of the reference spectrum:
\begin{equation}
\begin{split}
K_A =  \frac{\int_{f\in F_A} |TX_{ref}(f)|^2 df}{\int_{f \in F_{BOI}} |TX_{ref}(f)|^2 df}, \\
K_B =  \frac{\int_{f\in F_B} |TX_{ref}(f)|^2 df}{\int_{f \in F_{BOI}} |TX_{ref}(f)|^2 df},
\end{split}
\end{equation}
implying that $K_A + K_B = 1$, and in our current configuration: $K_A = K_B = \frac{1}{2}$.

The constant power constraint introduces dependencies in the NLN categories and they are not longer separable as in the previous section. The NSR can be decomposed into squared ($NSR_{NLNref,2}$), linear ($NSR_{NLNref,1}$), constant ($NSR_{NLNref,0}$) NSR contributions:
\begin{equation}
\begin{split}
NSR_{NLNref}^{(A)}= NSR_{NLNref,2}^{(A)} \Delta^2(A) +  NSR_{NLNref,1}^{(A)} \Delta(A) + \\    NSR_{NLNref,0}^{(A)},
\end{split}
\end{equation}	

Similarly, for the APSDs:
\begin{equation}
\begin{split}
\overline{S}_{NLNref}^{(N)} = \overline{S}_{NLNref,3}^{(N)}  \Delta^3(A) + \overline{S}_{NLNref,2}^{(N)} \Delta^2(A) + \\ \overline{S}_{NLNref,1}^{(N)} \Delta(A)  + \overline{S}_{NLNref,0}^{(N)} \, \, \, \, ,
\end{split}
\end{equation}	

For reference, Appendixes A, and B includes definition for APSD constant power, the existing symmetries on the terms, and discussions of certain properties of the fitting for this particular scenario. The maximum error observed was 0.6 dBs for both NLN and ASE categories.

Constant power perturbations do not change the total power launch into the fiber, which is beneficial for not affecting the mode of operation of deployed amplifiers and the overall communication link. 

\section{Simulations}
Table \ref{tab:simParameters} illustrates the parameters chosen for the split-step (SS) simulations. Non-Dispersion-Shifted Fiber (NDSF) was considered. To cover both intra- and inter-channel contributions, a single channel and a 3 channel configuration are considered in a 75 GHz grid. The middle channel has the shape of $F_{BOI}$ of Figure \ref{fig:PertSimplExampleTotal}, corresponding to two sub-carriers of 20 GBauds, while the other channels are Nyquist shaped of 50 GBauds. 

All neighboring channels have equal launched power 2 dBm. The power of $|TX_{ref}(f)|^2$ is the same as the neighboring channels, but the power of the perturbed spectrum is allowed to be different.

We used the spectrum of Figure \ref{fig:PertSimplExampleTotal} where the width of $F_A$ and $F_B$ is set to 20 GHz, while the width of $F_N$ is 10 GHz, for a occupancy of 50 GHz. The span length is 100km, up to 30 spans of fiber. Gaussian data mimics the behavior of a high cardinality modulation format and strengthens the accuracy of the GN model in use.

We include  enough perturbation realizations to invert $(\overline{\overline{\Delta}}^{T} \overline{\overline{\Delta}})^{-1}$. $\Delta(A)$ and $\Delta(B)$, were varied between -2 to 2 dB in steps of 1 dB for a total of 25 perturbation instances.

We normalize the signal such that their APSD of their BOI is equal to 1. E.g.: the PSD of a rectangular Nyquist shaped pulse would be equal to 1. We consider the same normalization for the different conditions of Table \ref{tab:simParameters}. 

Considering practical experimental limitations, the totality of $F_N$ cannot be integrated given the linewidth of existing Optical Scope. In this analysis only 85\% of the totality of the $F_N$ category is integrated, the edges of the region are discarded.

\begin{figure*}
\subfloat[$\Delta(A)=0$ dB, $\Delta(B)=0$ dB]{\includegraphics[scale=0.42]{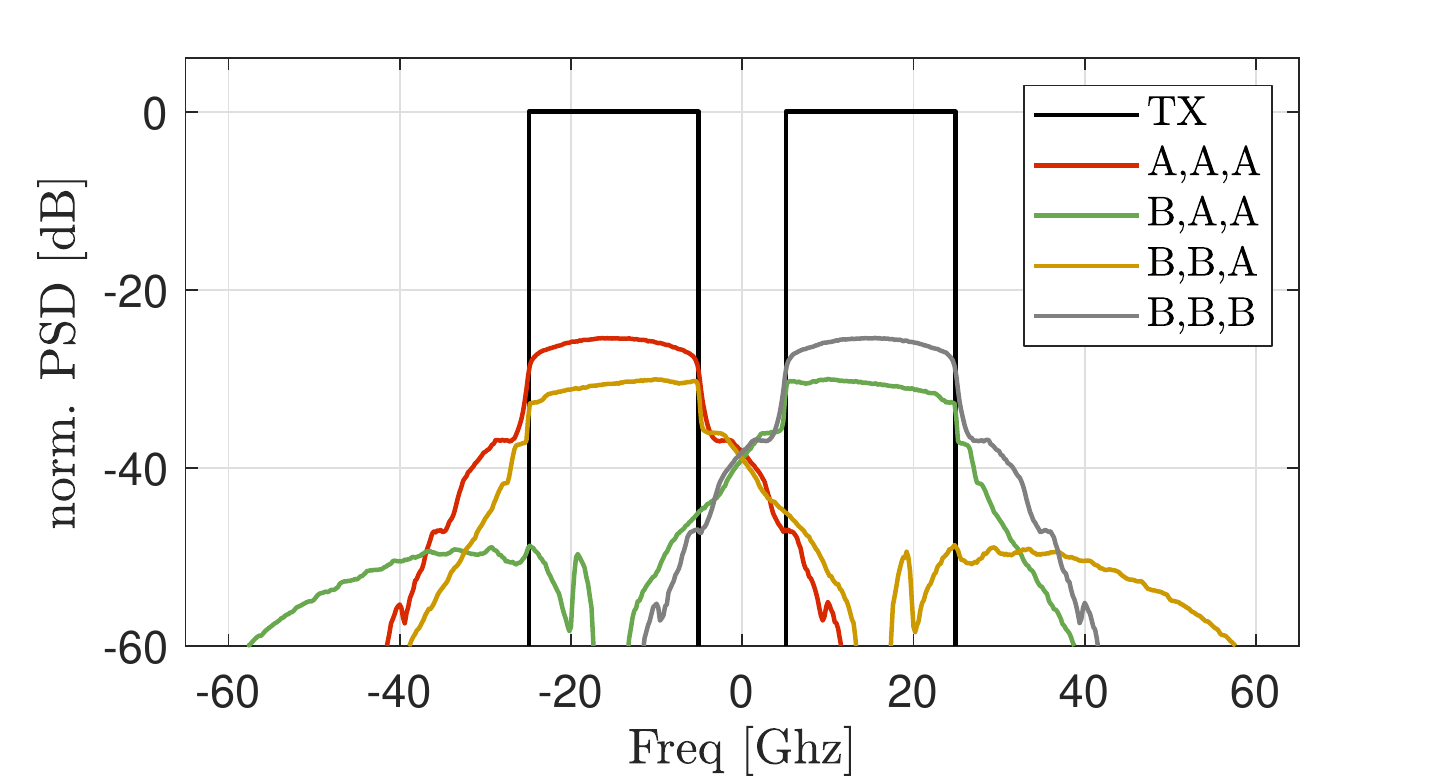}}
\subfloat[$\Delta(A)=5$ dB, $\Delta(B)=-5$ dB]{\includegraphics[scale=0.42]{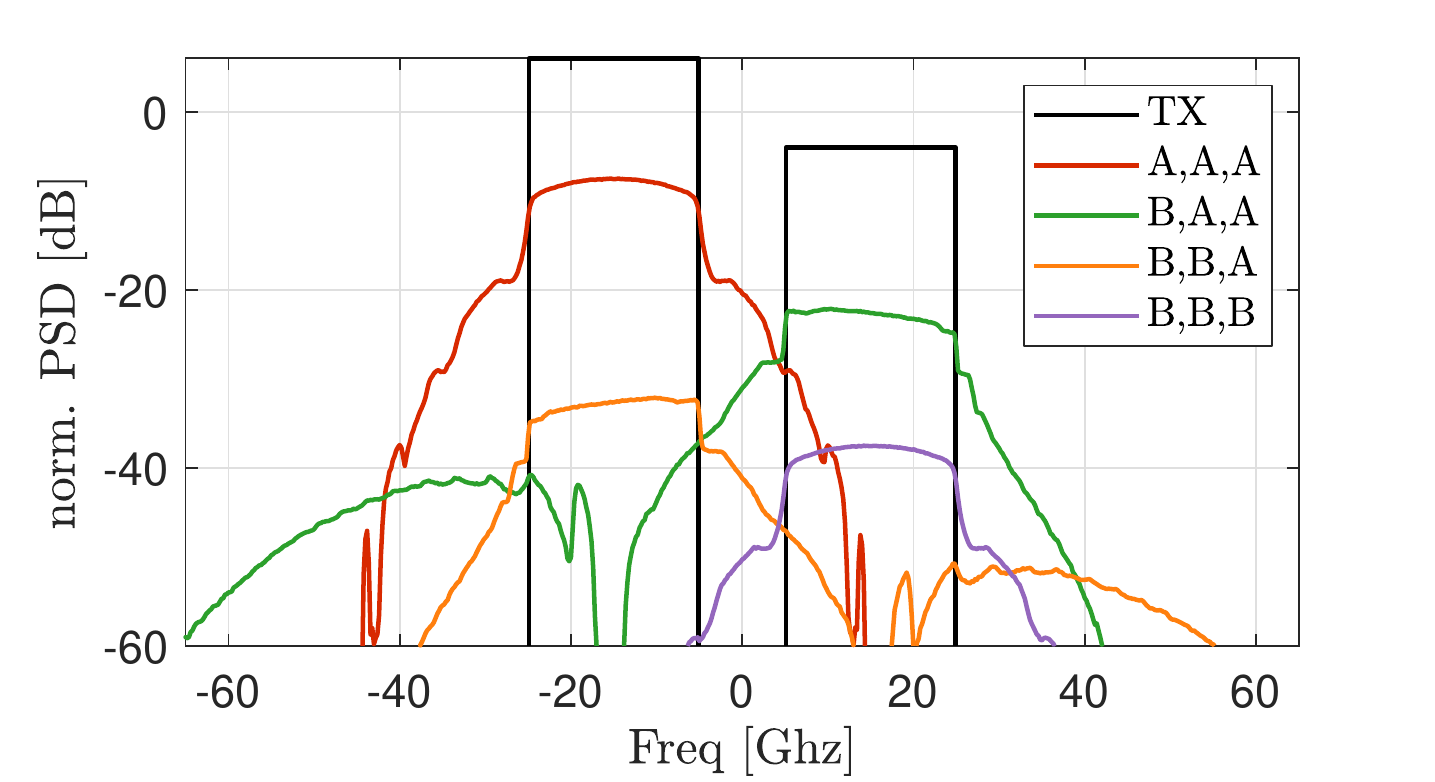}}
\subfloat[$\Delta(A)=-5$ dB, $\Delta(B)=5$ dB]{\includegraphics[scale=0.42]{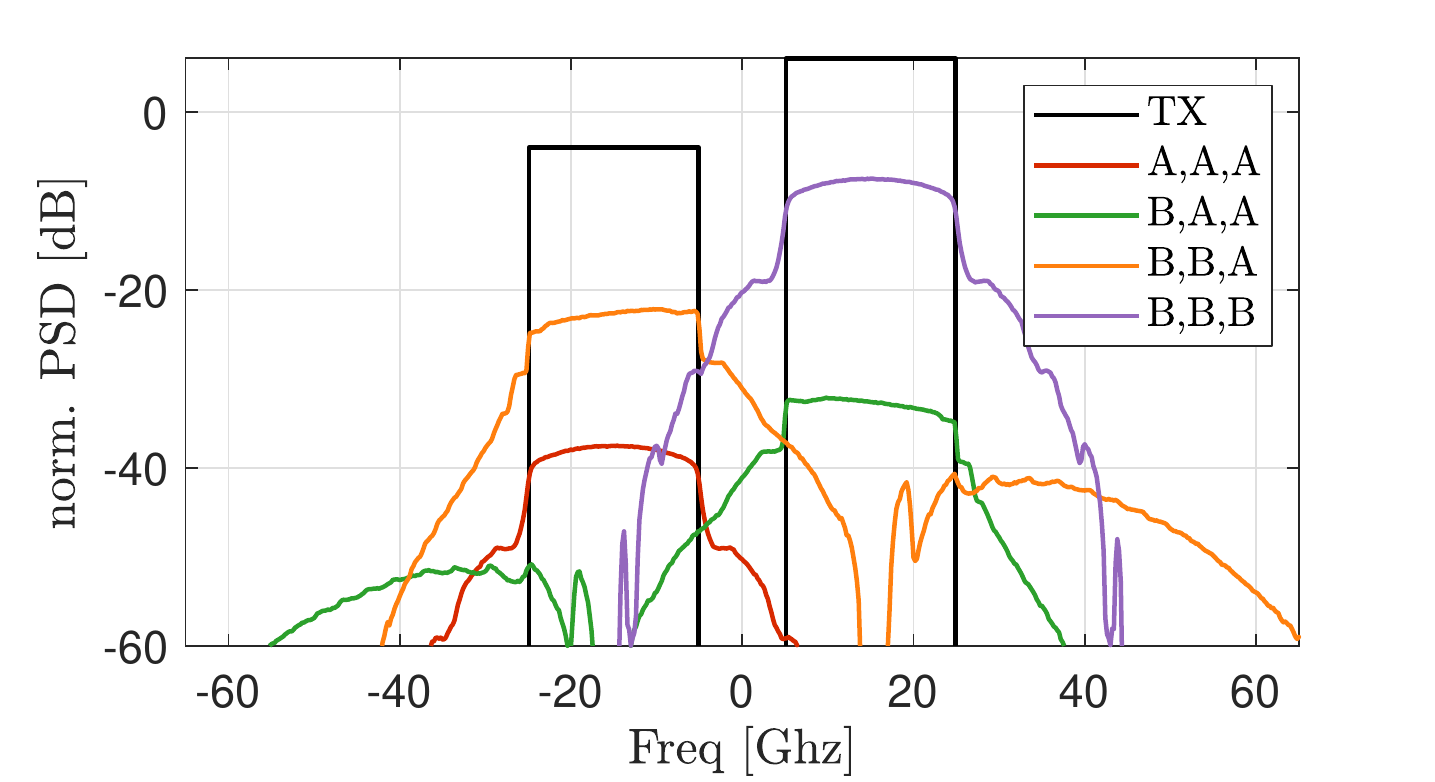}}
\\
\subfloat[$\Delta(A)=0$ dB, $\Delta(B)=0$ dB]{\includegraphics[scale=0.42]{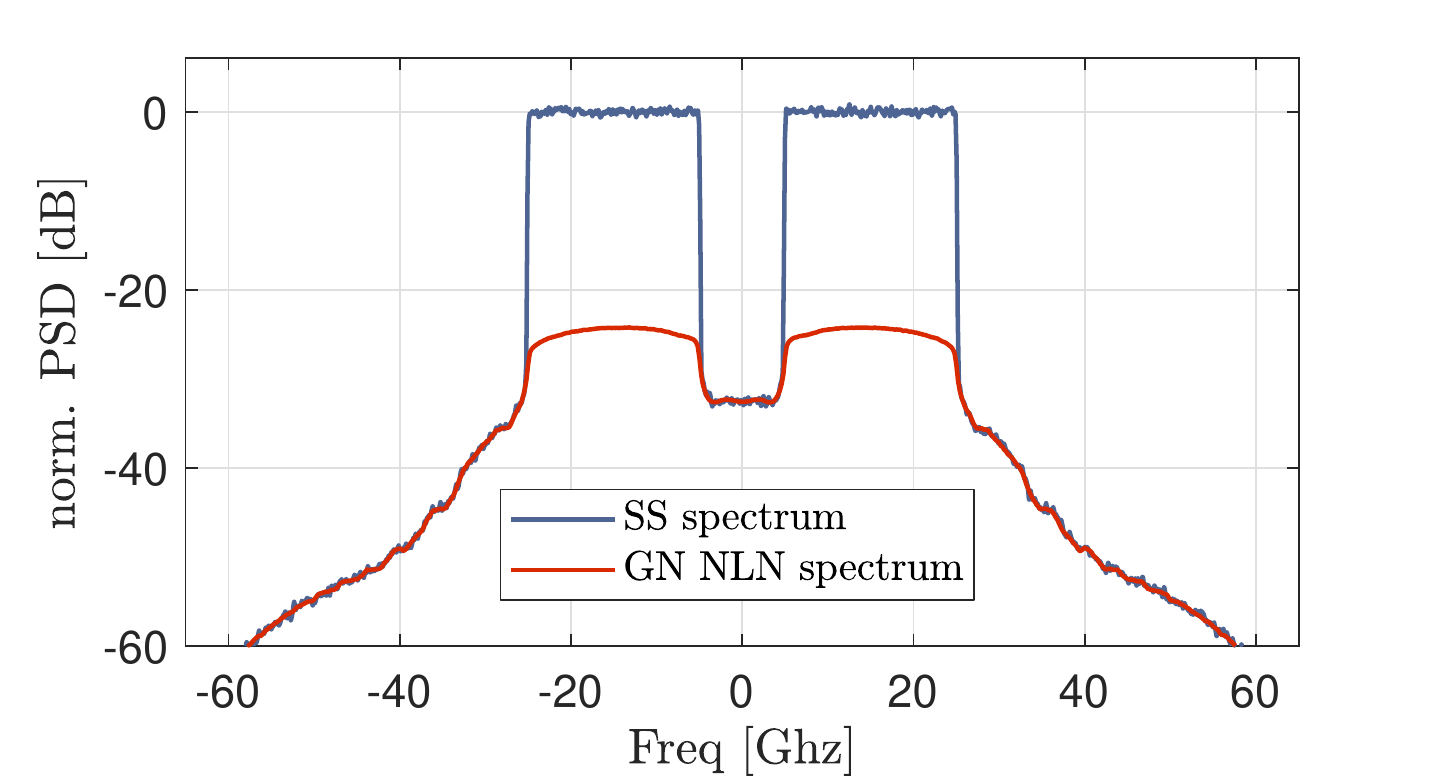}}
\subfloat[$\Delta(A)=5$ dB, $\Delta(B)=-5$ dB]{\includegraphics[scale=0.42]{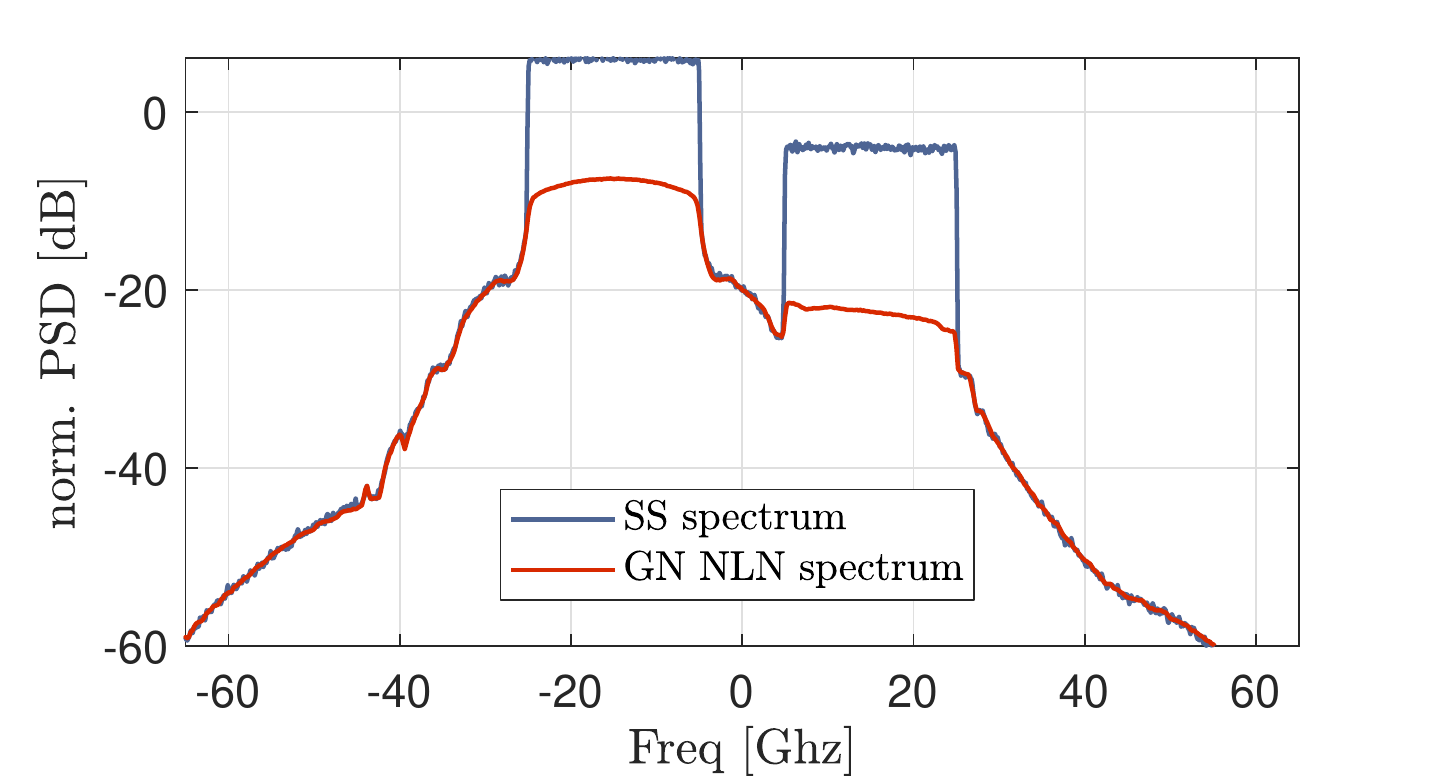}}
\subfloat[$\Delta(A)=-5$ dB, $\Delta(B)=5$ dB]{\includegraphics[scale=0.42]{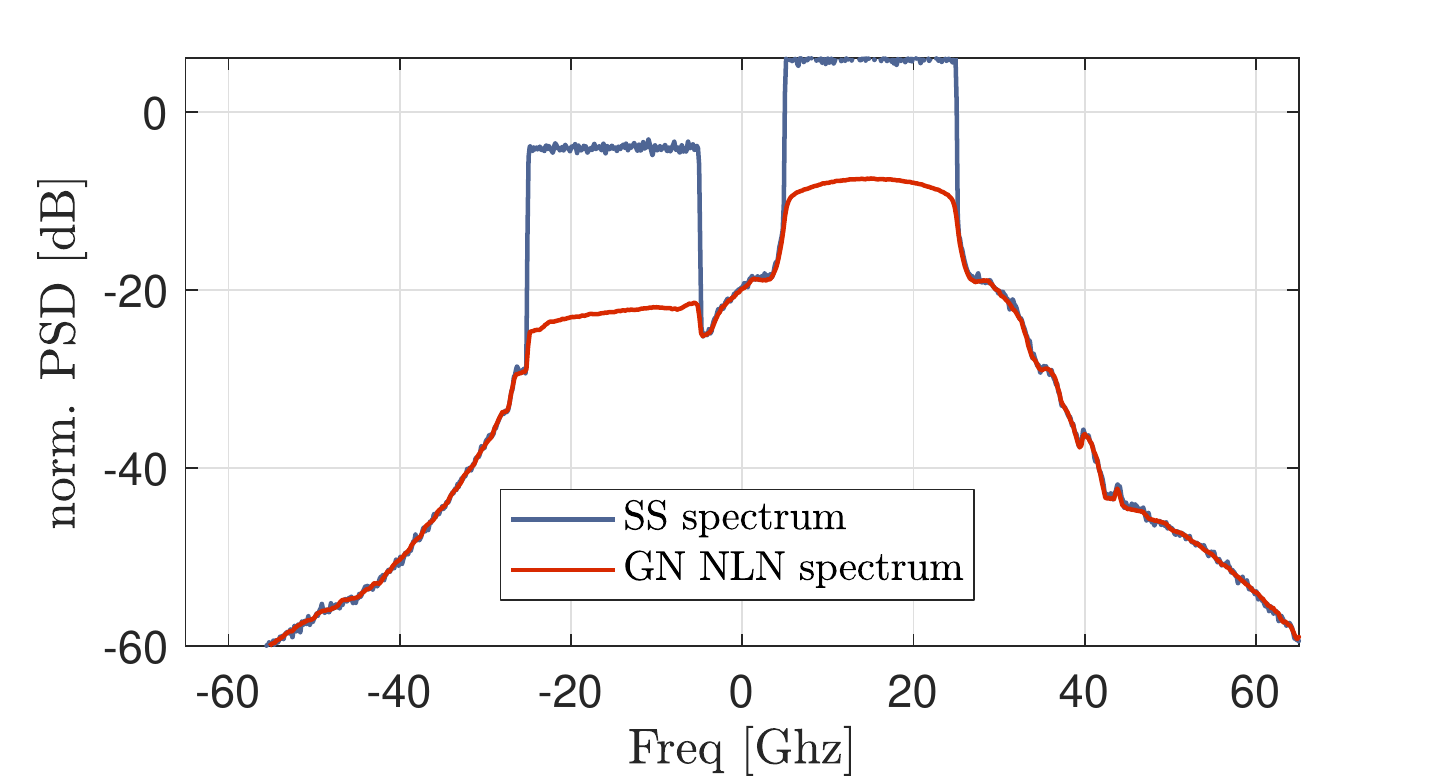}}
\caption{$|NLN_{intra}(f)|^2$ categories and PSDs for conditions of Table \ref{tab:Parameters}, 10 spans of NDSF, 2dBm.}
\label{fig:NLNtermsExpIntra}
\end{figure*}

\begin{figure*}
\subfloat[$\Delta(A)=0$ dB, $\Delta(B)=0$ dB]{\includegraphics[scale=0.42]{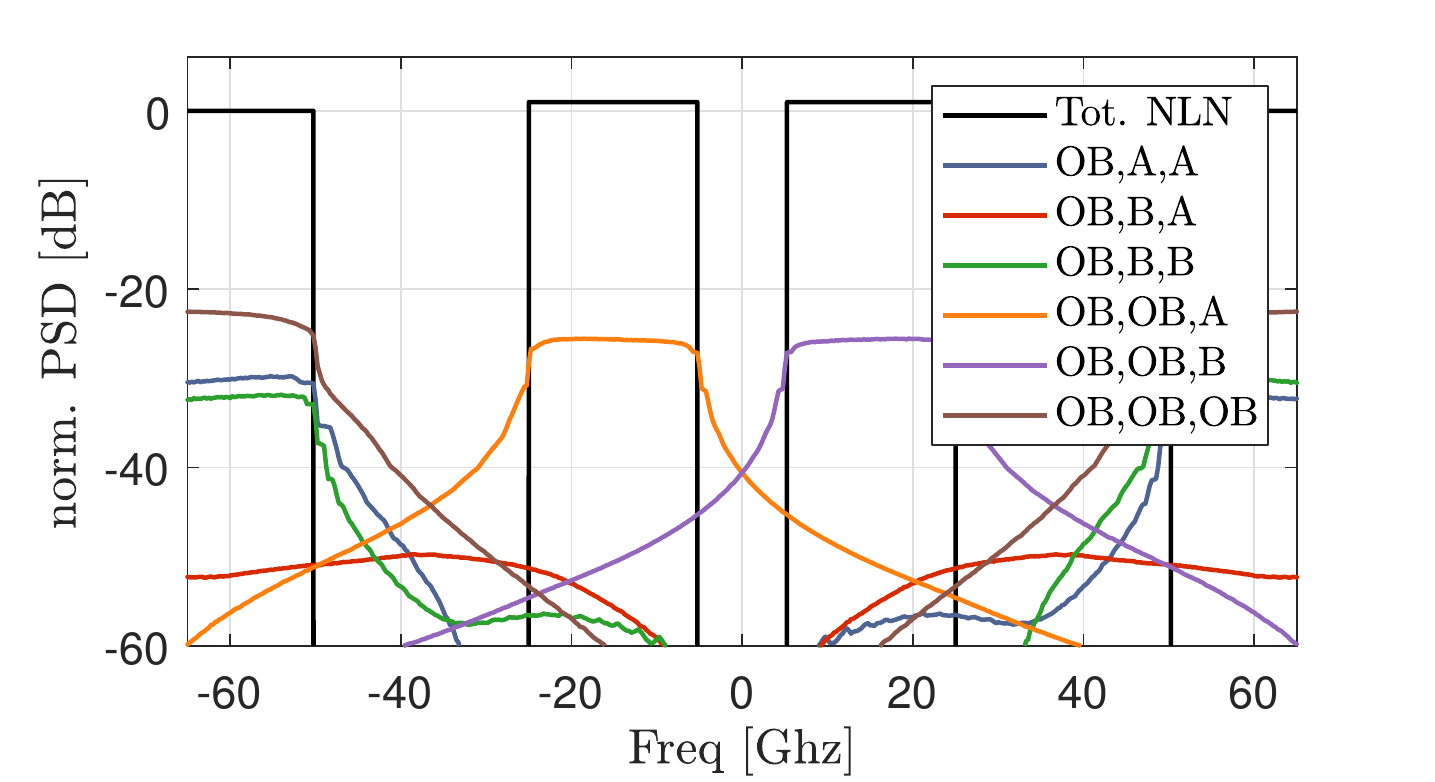}}
\subfloat[$\Delta(A)=5$ dB, $\Delta(B)=-5$ dB]{\includegraphics[scale=0.42]{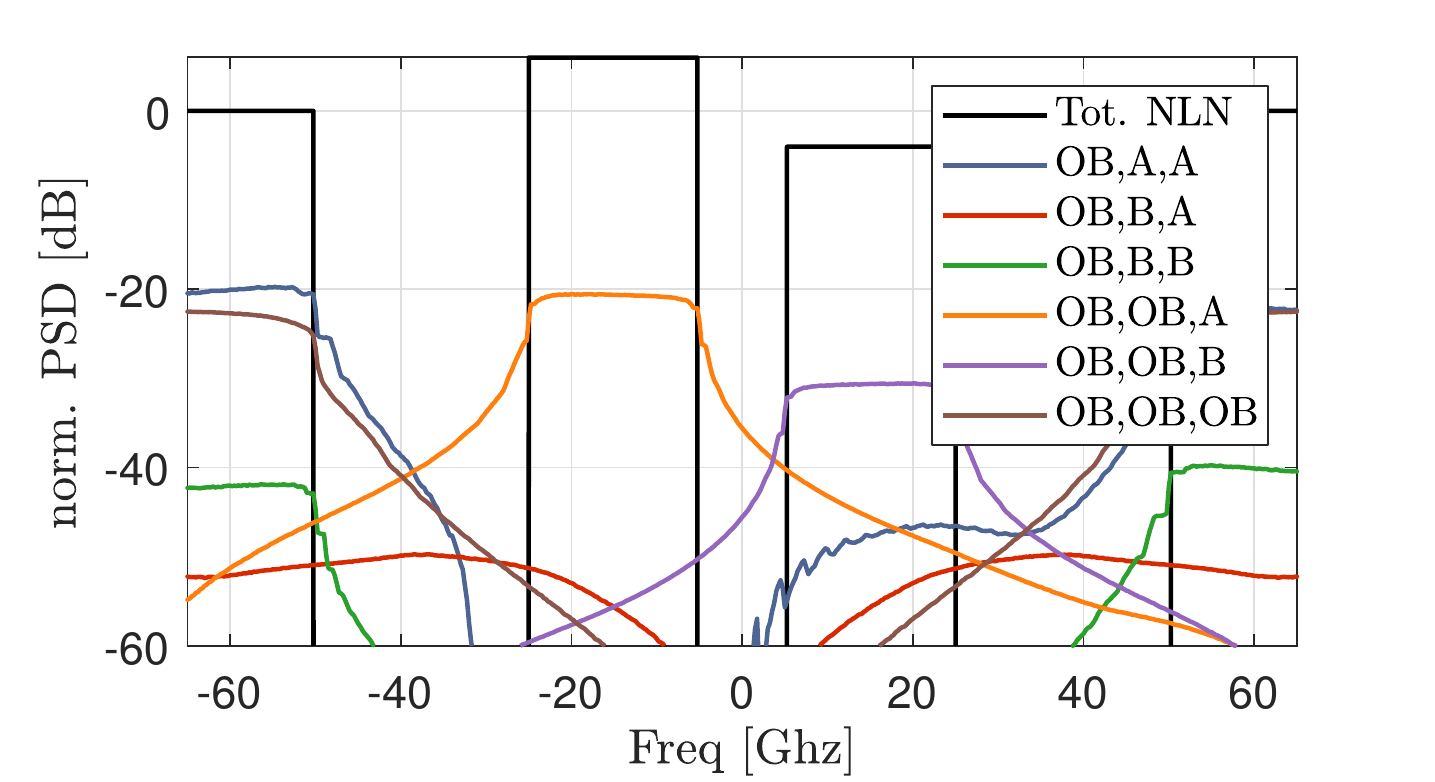}}
\subfloat[$\Delta(A)=-5$ dB, $\Delta(B)=5$ dB]{\includegraphics[scale=0.42]{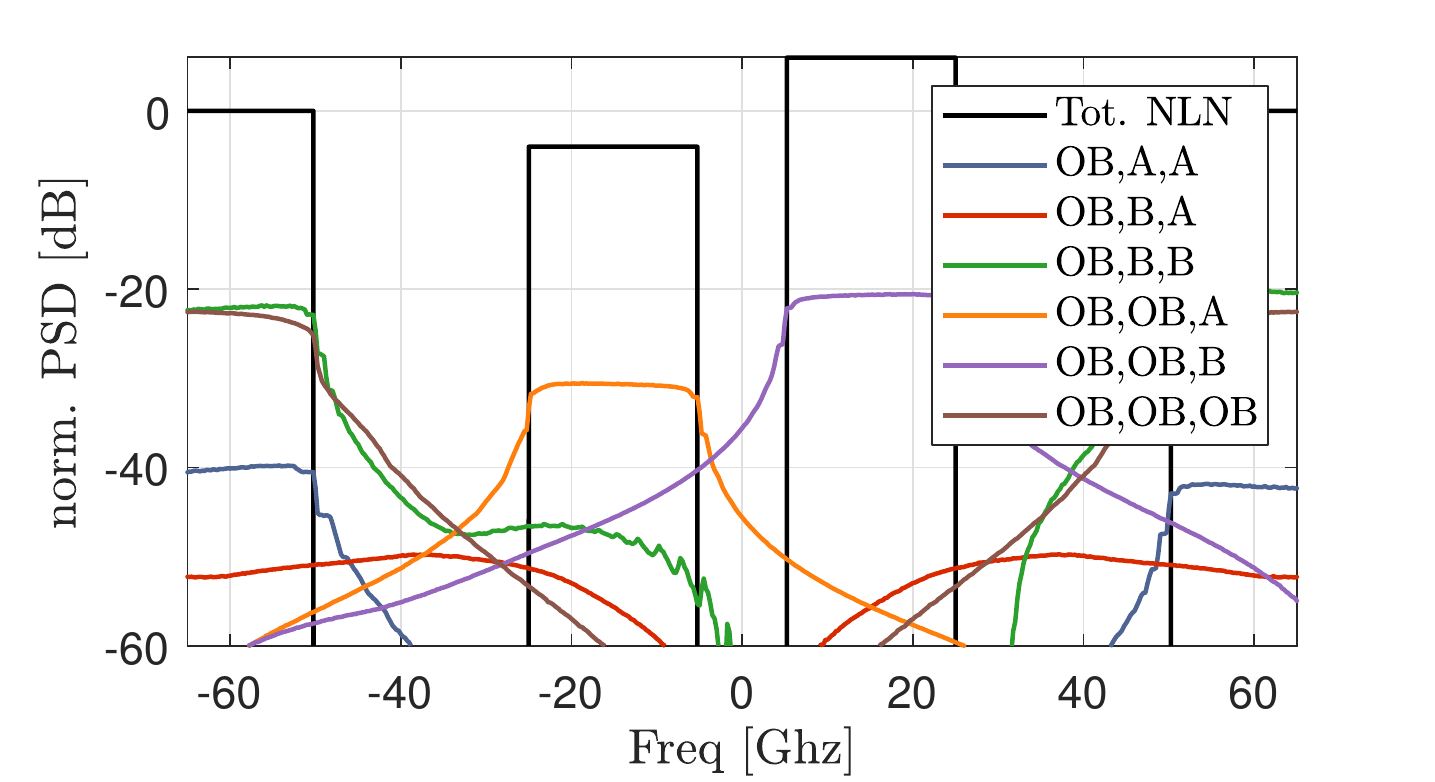}}
\\
\subfloat[$\Delta(A)=0$ dB, $\Delta(B)=0$ dB]{\includegraphics[scale=0.42]{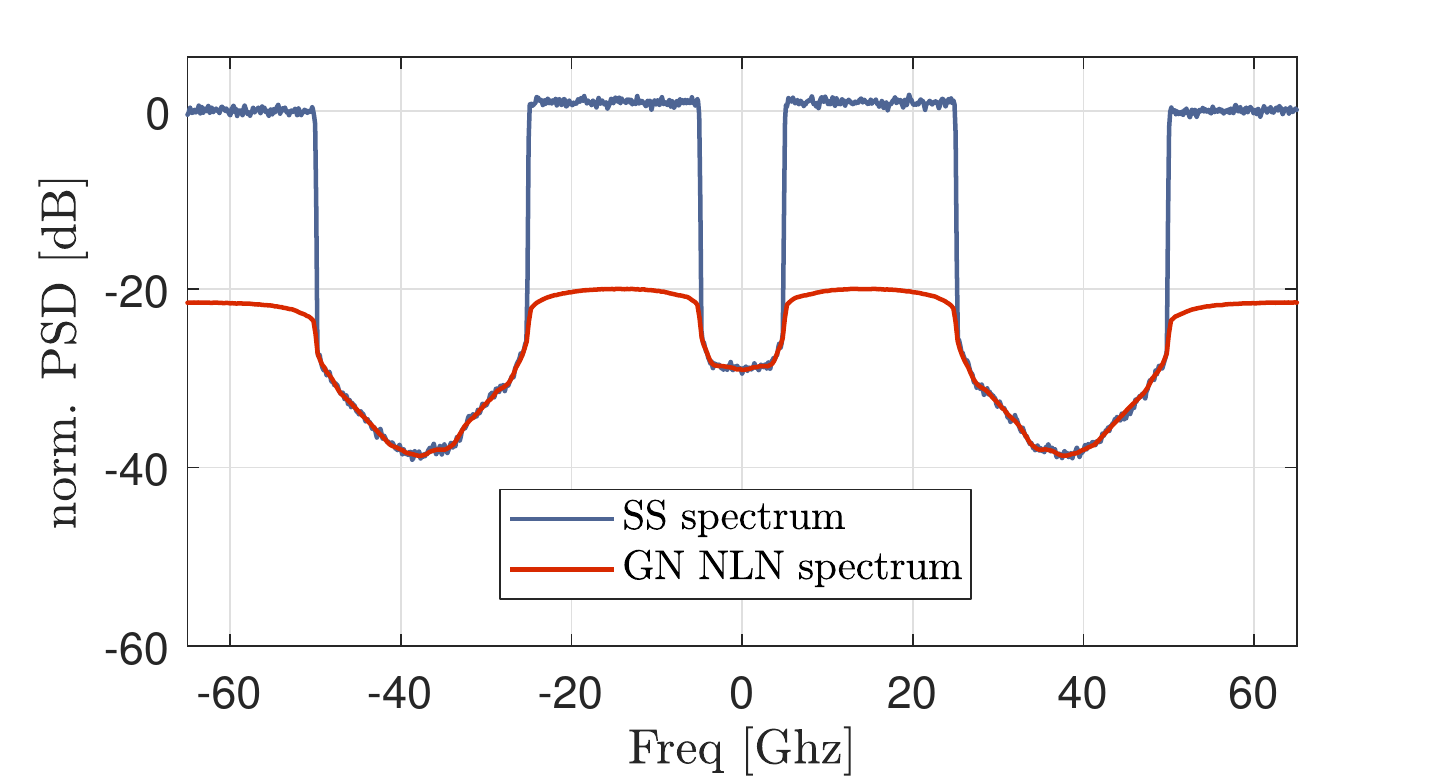}}
\subfloat[$\Delta(A)=5$ dB, $\Delta(B)=-5$ dB]{\includegraphics[scale=0.42]{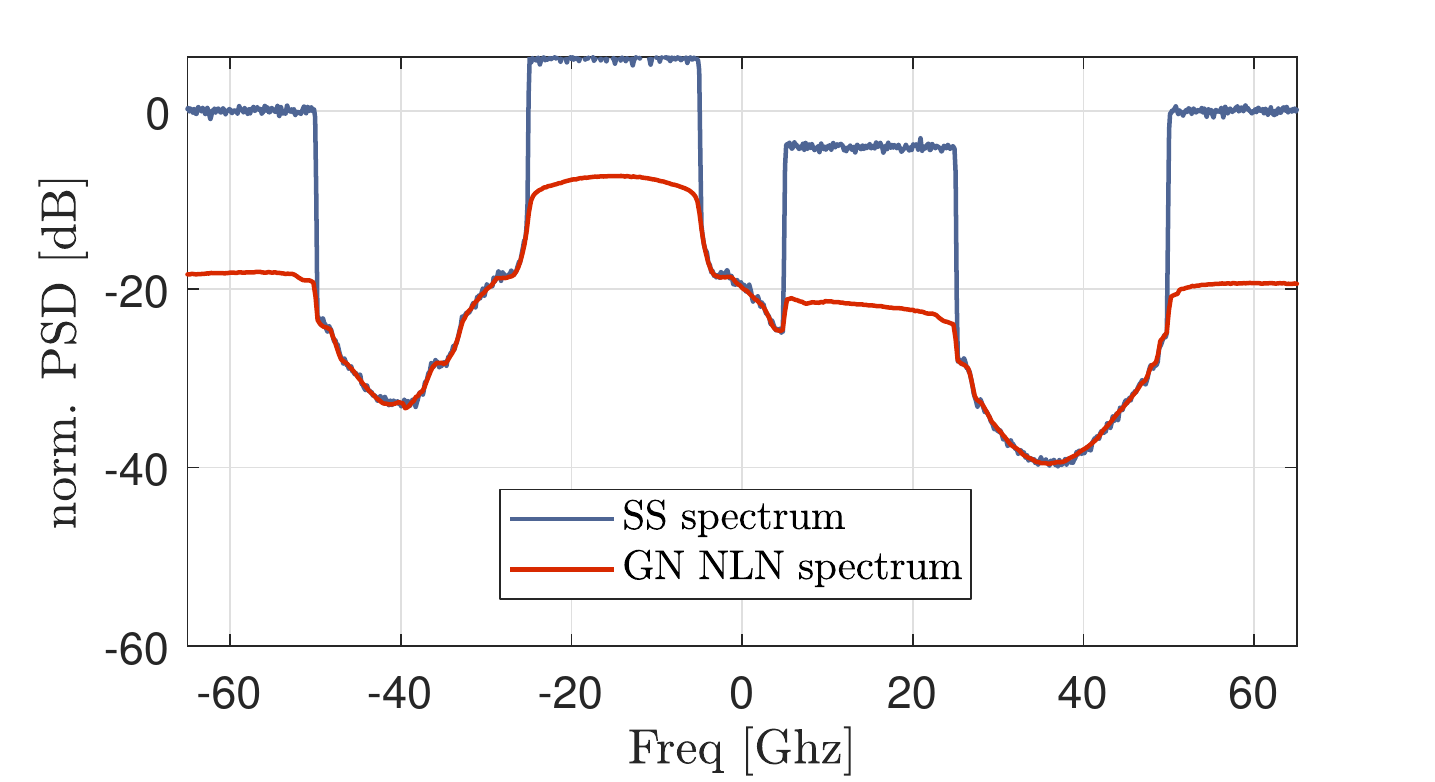}}
\subfloat[$\Delta(A)=-5$ dB, $\Delta(B)=5$ dB]{\includegraphics[scale=0.42]{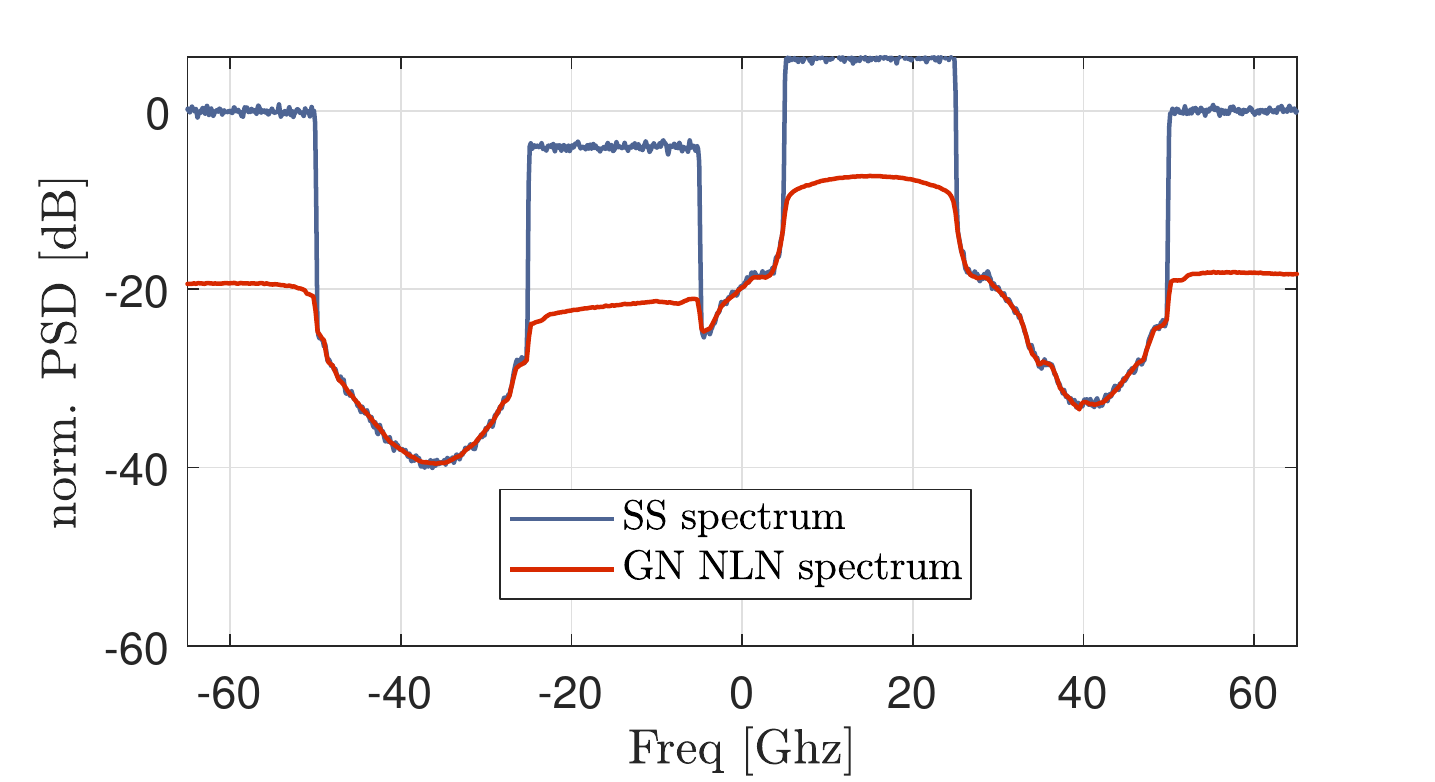}}
\caption{$|NLN_{inter}(f)|^2$ categories and PSDs for conditions of Table \ref{tab:Parameters}, 10 spans of NDSF, 2dBm.}
\label{fig:NLNtermsExpInter}
\end{figure*}

Figure \ref{fig:NLNtermsExpIntra} (d,e,f), and Figure \ref{fig:NLNtermsExpInter} (d,e,f), compares the spectral of the GN and the SS methods. The SS method plots the totality of the spectrum given that the underling NLN under the signal is not directly measurable, while the GN analysis only accounts for the NLN component. Additionally, Figure \ref{fig:NLNtermsExpIntra} (a,b,c), and Figure \ref{fig:NLNtermsExpInter} (a,b,c) plots the NLN categories for the counterparts of Figure \ref{fig:NLNtermsExpIntra} (d,e,f), and Figure \ref{fig:NLNtermsExpInter} (d,e,f).

\begin{table}[htp] 
   \centering
   \small
\caption{Parameters for the Simulation Verification.
} \label{tab:Parameters}
\begin{tabular}{l | r  r r}
 \centering%
Parameter &	\multicolumn{2}{c}{Value} \\
 \hline	
Fiber type							&	\multicolumn{2}{c}{NDSF}	\\
\hline
Nonlinear Coeff. ($\gamma$) [1/W/km]&	\multicolumn{2}{c}{1.3}		\\
\hline
Dispersion Parameter (D) [ps/nm/km]	&	\multicolumn{2}{c}{16.7}		\\
\hline
Attenuation ($\alpha$) [dB/km]		&	\multicolumn{2}{c}{0.2}		\\
\hline
num. of WDM channels (Nch)			&	\multicolumn{2}{c}{1, 3}		\\
\hline
Channel Spacing [GHz]				&	\multicolumn{2}{c}{75}		\\
\hline
CH Launch power	[dBm]				&	\multicolumn{2}{c}{2}		\\
\hline
Modulation Format					& 	\multicolumn{2}{c}{Gaussian}\\
 \hline
Span length [km]					&	\multicolumn{2}{c}{100}			\\
 \hline
Number of Spans						&	\multicolumn{2}{c}{1:30}		\\
 \hline
Width of $F_A$, $F_B$ [GHz]			&	\multicolumn{2}{c}{20}		\\
\hline
Width of $F_N$ [GHz] 				&	\multicolumn{2}{c}{10}		\\
\hline
$\Delta_k(A)$	[dB]				&	\multicolumn{2}{c}{-2:1:2}		\\
\hline
$\Delta_k(B)$	[dB]				&	\multicolumn{2}{c}{-2:1:2}		\\
\hline
Out-of-band signal 	shape		    &	Nyquist	\\
\hline
Amplifier N.F	[dB]	            &	\multicolumn{2}{c}{4.5}	\\
\hline
Gain [dB]						    &	\multicolumn{2}{c}{20}	\\
\hline
$F_{BOI}$  [GHz]                    &	\multicolumn{2}{c}{50}	\\
\hline
$F_{A}$  [GHz]                      &	\multicolumn{2}{c}{20}	\\
\hline
$F_{B}$  [GHz]                      &	\multicolumn{2}{c}{20}	\\
\hline
$F_{N}$  [GHz]                      &	\multicolumn{2}{c}{10}	\\
\hline
Symb. Rate (BOI sub-carrier)[GBaud] &	\multicolumn{2}{c}{20}	\\
\hline
Symb. Rate (out-of-band)[GBaud]     &	\multicolumn{2}{c}{50}	\\
\hline
\end{tabular}
\label{tab:simParameters}
\end{table}%

\begin{figure*}
\centering
\subfloat[Intra-channel APSD categories from GN and noise-less SS for single channel]{\includegraphics[scale=0.6]{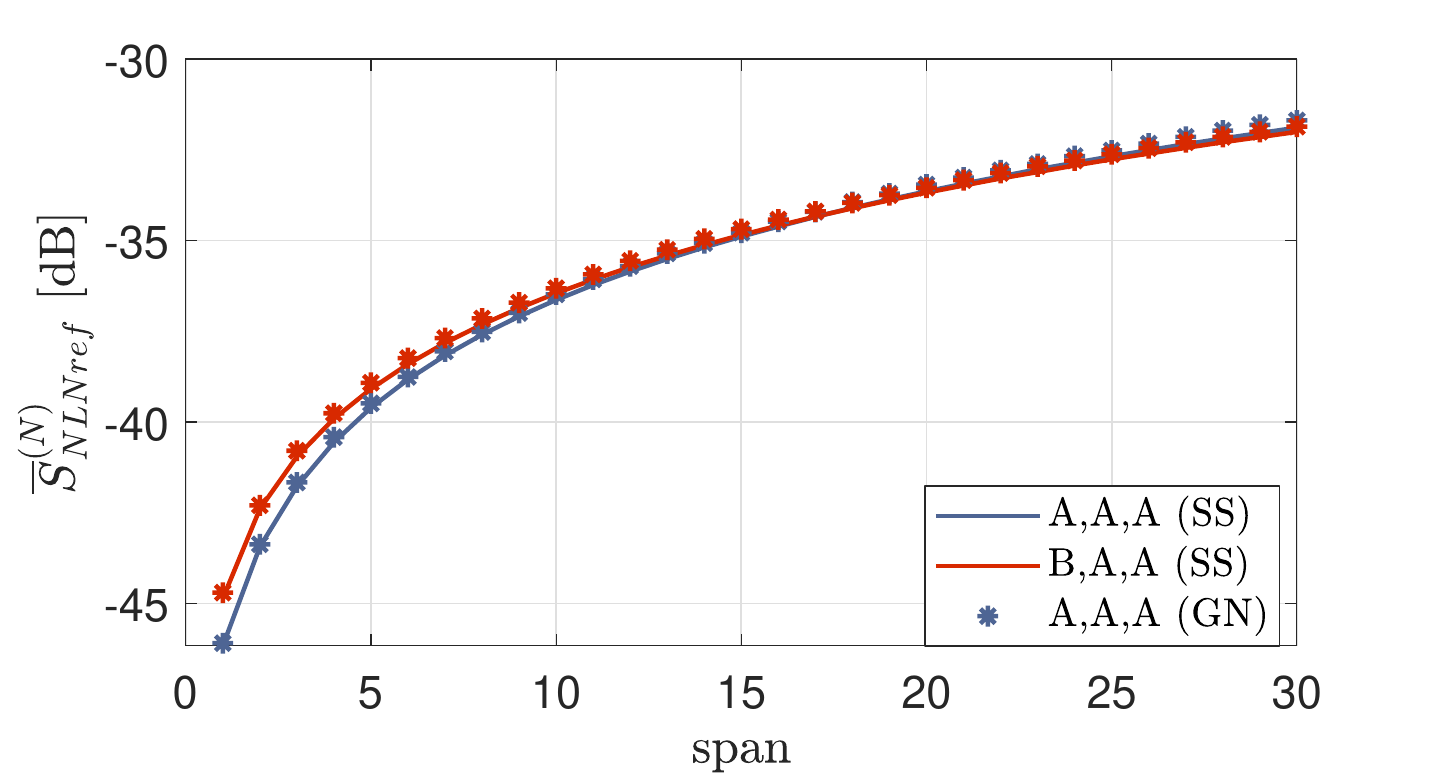}}
\subfloat[APSD categories in $F_N$]{\includegraphics[scale=0.6]{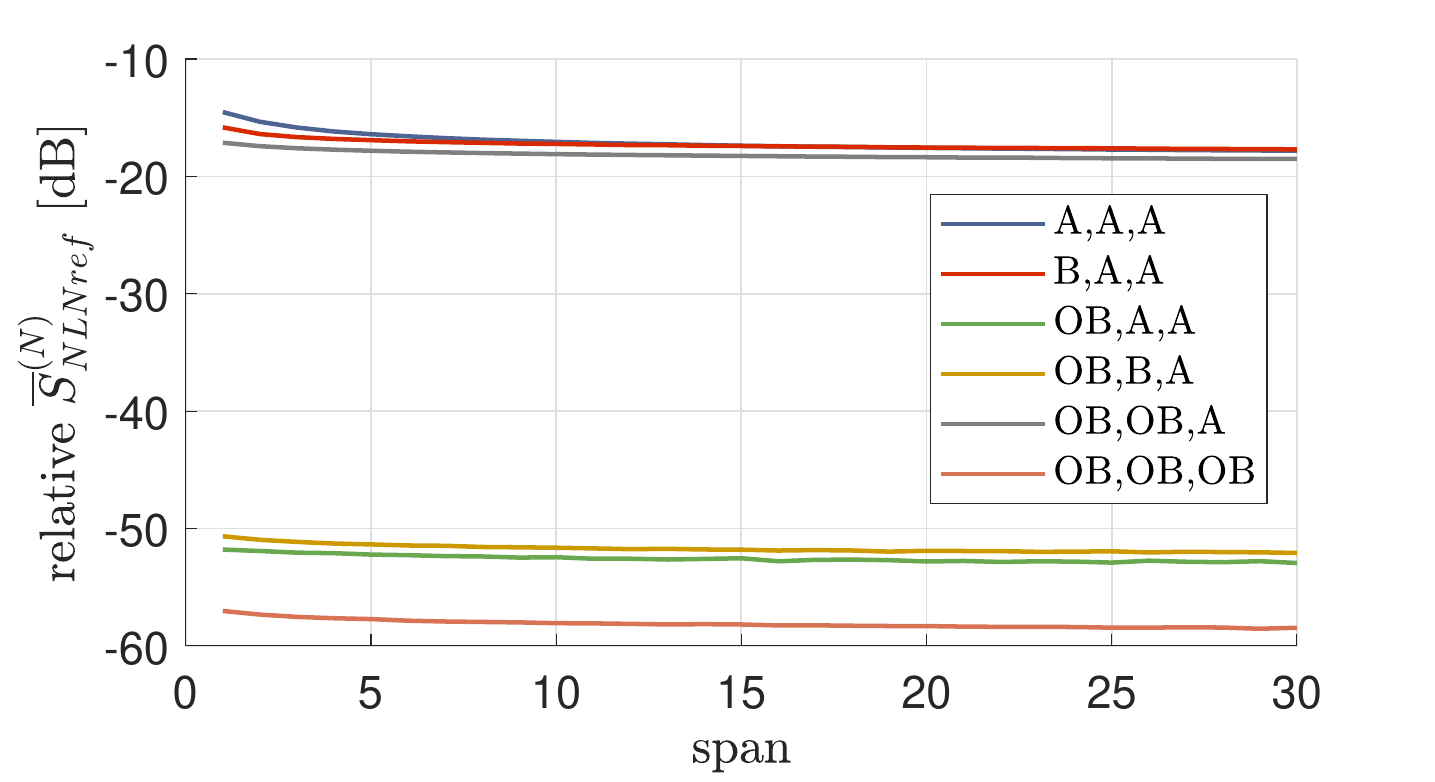}}
\\
\subfloat[Inter-channel APSD categories from GN and noise-less SS for 3 channel scenario]{\includegraphics[scale=0.6]{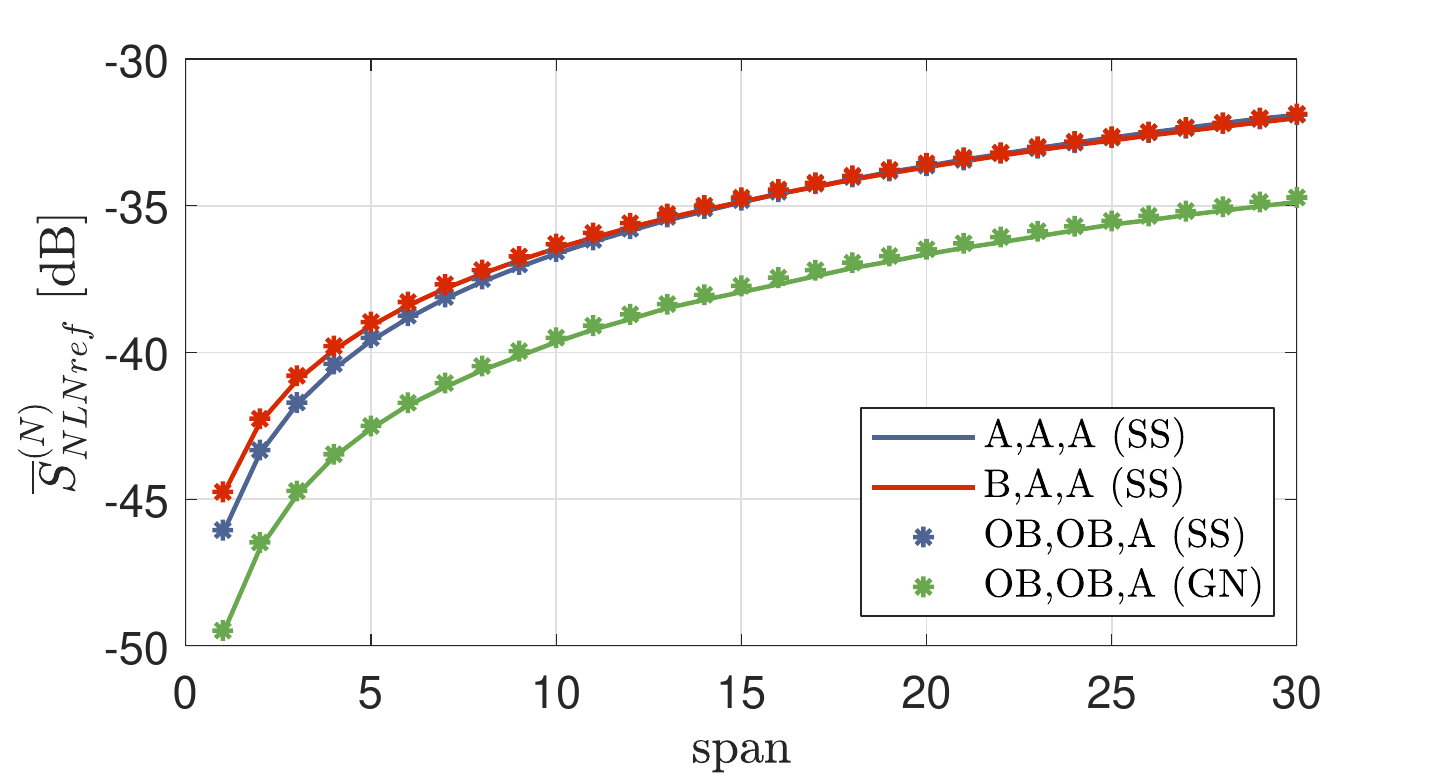}}
\subfloat[Intra-channel NSR categories]{\includegraphics[scale=0.6]{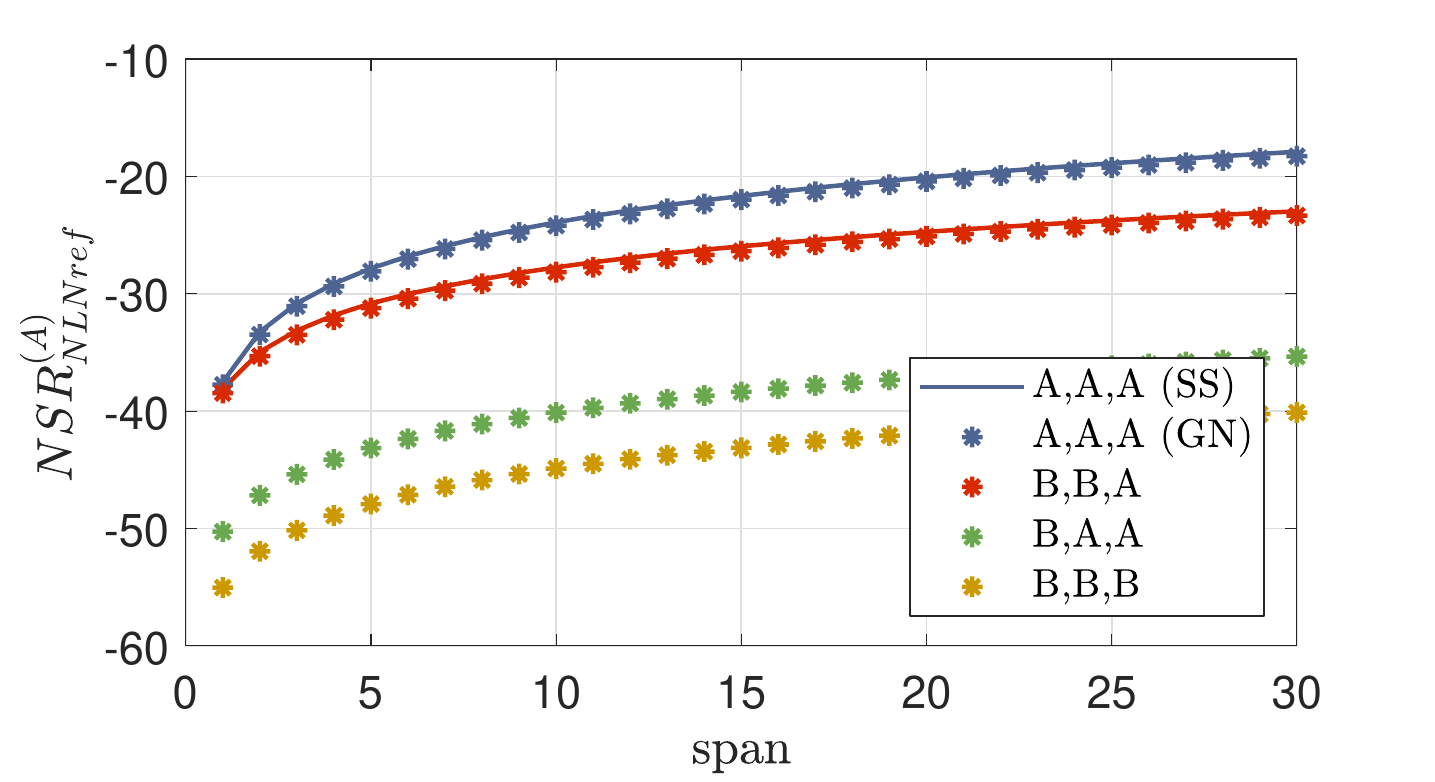}}
\caption{NLN fitted categories for the conditions of Table \ref{tab:Parameters}.}
\label{fig:NLNtermsFitting}
\end{figure*}

\subsection{APSD Single Channel Fitting}
$\overline{S}_{NLN_{ref}}$ was calculated for the received normalized PSD after the transmitted spans. By applying Equation \ref{eq:FittingCase}, it was possible to compare the GN and SS results. Figure \ref{fig:NLNtermsFitting} (a) illustrates the agreement between SS and GN for the [A,A,A] and [B,B,A] categories, continuous lines corresponds to the split-step simulations, whereas start dots corresponds to the GN calculations. 

By symmetries of Equation \ref{eq:IntraNLNsymFragments}, [B,B,B] and [B,B,A] were not included given their value is identical to [A,A,A] and [B,A,A], respectively.

\subsection{APSD WDM Channel Fitting}
Figure \ref{fig:NLNtermsFitting} (b) illustrates the relative sizes of noise APSD terms $F_N$, where the symmetric terms of Equations 
\ref{eq:IntraNLNsymFragments} and \ref{eq:InterNLNsymFragments}, are omitted for simplicity.

From Figure \ref{fig:NLNtermsFitting} (b), it is observable that the dominant contributions are intra-channel categories, and inter-channel categories corresponding to XPM: [OB,OB,A] and [OB,OB,B], this amounts to approximating the totality of NLN as intra-channel and XPM terms. 

Given the constant nature of the [OB,OB,OB] category, it is not separable from the ASE contribution. Figure \ref{fig:NLNtermsFitting} (b)  shows that the term of the [OB,OB,OB] category can be consider negligible (-60 dB) compared to the ASE contribution (typically around -30 to -5 dB). 

Figure \ref{fig:NLNtermsFitting} (c) illustrates the results of the fitting for intra-channel and inter-channel contributions, confirming that Intra-channel contributions are the same as in the case of a single channel transmission of Figure \ref{fig:NLNtermsFitting} (a). Additionally, the XPM category of [OB,OB,A] is also plotted showing a good agreement between GN and SS. 

Because of symmetries, the [OB,OB,B] category being equal to [OB,OB,A] is not included on the plot.

\subsection{NSR Single Channel Fitting}
The verification of the NSR fitting was performed for the intra-channel components only, performed with two 20 GHz sub-carriers separated by a 10 GHz gap of Figure \ref{fig:PertSimplExampleTotal}. 

Figure \ref{fig:NLNtermsFitting} (d) corresponds to a 2 dBm  launch power signal, which is approximately the optimum power for this scenario. For our fitting, the NSR of [B,B,B] and [B,A,A] categories are considered negligible, this is consistent with the XPM approximation. 

\section{Experimental Verification }
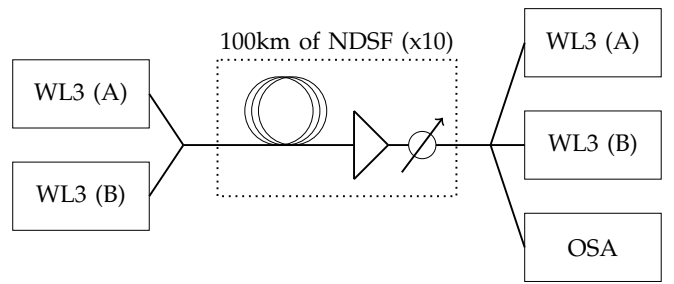
\begin{figure}[h!]
\centering
\resizebox{1\linewidth}{!}
{
\begin{tikzpicture}
\newcommand\AMP[4]{
\draw [thick] (#1,-#4+#2)	--	(#1,#4+#2);
\draw [thick] (#1, #4+#2)	--	(#1+#3,#2);
\draw [thick] (#1,-#4+#2)	--	(#1+#3,#2);
}
\newcommand\fiber[4]{
\draw (#1+#4,#2+#3) circle (#3);
\draw (#1,#2+#3) 	circle (#3);
\draw (#1-#4,#2+#3) circle (#3);
}
\draw [thick,-] (1,0)	--	(3.5,0);
\AMP{3.5}{0}{0.5}{0.5};
\draw [thick,-] 	(4,0)		--	(4.3,0);
\draw (4.5,0) circle (0.2);
\draw [thick,-] 	(4.7,0)		--	(5.5,0);
\draw [thick,->] 	(4.2,-0.4)     --	(4.8,0.4);	

\fiber{2.5}{0}{0.5}{0.1};
\draw[draw=black] (-1-0.5,0.5+0.75) rectangle node{WL3 (A)} (1-0.5,-0.5+0.75);
\draw[draw=black] (-1-0.5,0.5-0.75) rectangle node{WL3 (B)} (1-0.5,-0.5-0.75);

\draw [thick,-] (1-0.5,0.75)	--	(1,0);
\draw [thick,-] (1-0.5,-0.75)	--	(1,0);

\draw[draw=black,dotted,thick](1.5,1.25) rectangle (5,-0.75);
\draw 			(3.25,1.5) node {100km of NDSF (x10)};

\draw[draw=black,fill=white] (5.5+0.5,0.5+1.5) rectangle node{WL3 (A)} 		(7.5+0.5,-0.5+1.5);
\draw[draw=black,fill=white] (5.5+0.5,0.5) rectangle node{WL3 (B)} 			(7.5+0.5,-0.5);
\draw[draw=black,fill=white] (5.5+0.5,0.5-1.5) rectangle node{OSA} 			(7.5+0.5,-0.5-1.5);

\draw [thick,-] (5.5,0)	--	(6,1.5);
\draw [thick,-] (5.5,0)	--	(6,0);
\draw [thick,-] (5.5,0)	--	(6,-1.5);

\end{tikzpicture}
}
\caption{Diagram of the experimental setup}
\label{fig:ExperimentalDiagram1}
\end{figure}

An experimental setup was built with 2 WaveLogic 3 (WL3) transceivers: WL3(A) and WL3(B), transmitting a 35GBaud QPSK signal in a 50GHz grid with 30000 ps of pre-compensated chromatic dispersion. The light path was 10 spans of 100 km of Non-Dispersion-Shifted Fiber (NDSF).  Figure \ref{fig:ExperimentalDiagram1} illustrates the experimental setup.

The launch power was varied as defined by $\Delta(A)$, $\Delta(B)$, in the range of -2 to 2 dB in steps of 2dB. The launch power of each channel was 4 dBm. The NSR was obtained from the reported BER from the card, and the transceiver contribution was subtracted to reduce the number of noise categories to estimate.

Given that applied perturbations disrupt the input power into the amplifiers, the amplifiers´s gain was kept constant but the launch power was corrected by a Variable Optical Atenuator (VOA). The input power into each span is set by an OSA.

\begin{figure*}[h]
\centering
\includegraphics[scale=0.6]{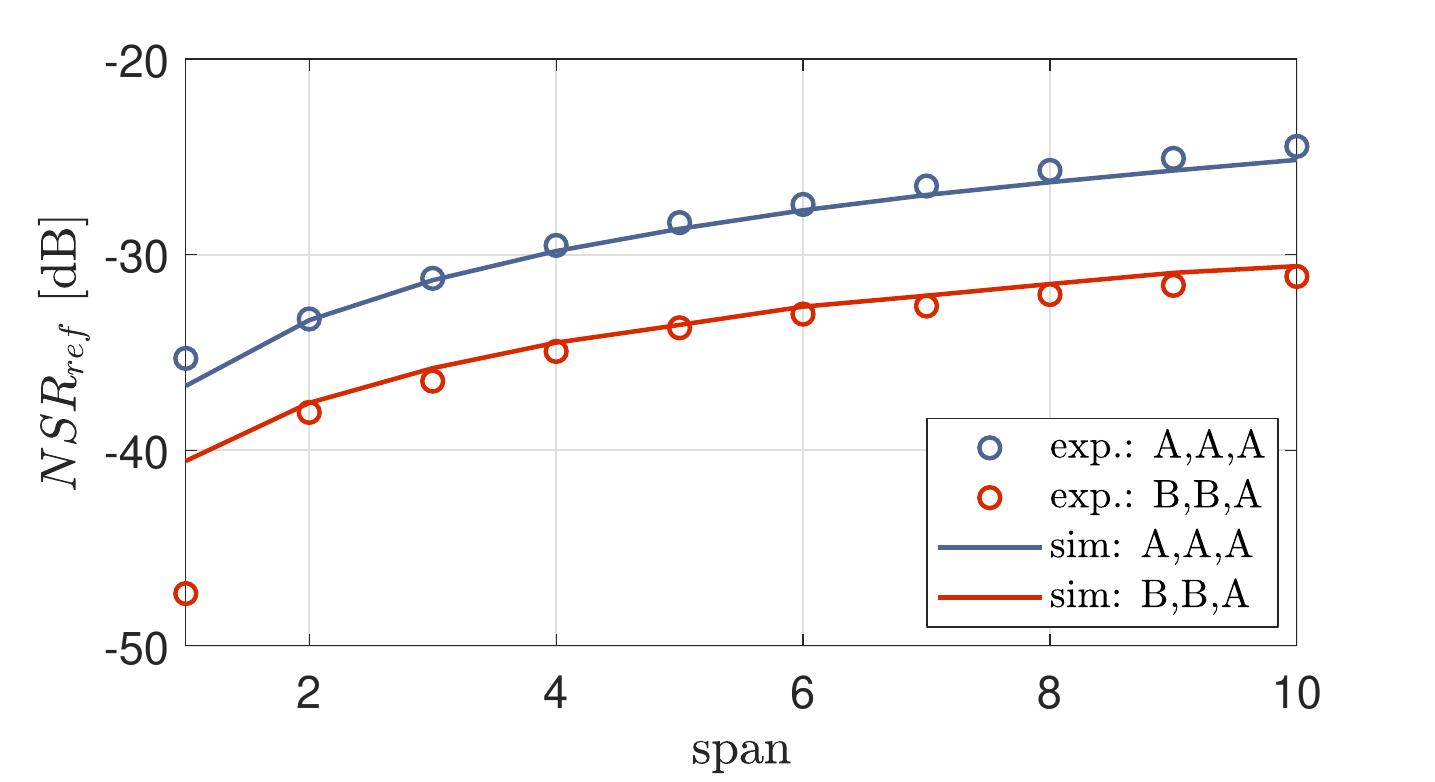}
\caption{Experimental Results}
\label{fig:ExpResults}
\end{figure*}

Figure \ref{fig:ExpResults} illustrates the experimental results, Figure \ref{fig:ExpResults} (a) shows the NLN NSR fitted as a function of the span count and compared with split-step simulations. The first spans are noisier since the NLN contribution is very small, good agreement was observed between both card NLN estimates and split-step estimates. 



\section{Conclusion and Future Work}
In this paper, we introduce a perturbation approach for the separation of optical noises. Our approach is based on the Gaussian Noise model and it allows to divide the nonlinear noise into categories that can be estimated by the introduction of a series of perturbations on the system.

We introduced a mathematical notation to describe noises from a average power spectral density and an NSR point of view, as well as a constant and variable power perturbation scenarios. We evaluate the agreement between split-step and GN models, and compare noiseless estimates of the GN and the split step for the case of variable power perturbation scenario. 

Finally, a small experimental setup is also performed with 2 WL3 transmitters. Good agreement was observed within simulations and experiments for the NLN components, the ASE terms were also estimated with a reasonable agreement with the ASE estimates from an OSA.

Given the current trend of optical transceivers towards subcarrier transmission, our technique is relevant and suitable for implementation in current and next generation transceivers. It is capable of decomposing the SNR into several ASE and NLN contributions differentiating between intra-channel NLN, inter-channel NLN, and ASE terms.

As future work, transmission with higher number of WDM channels as well as constant power constraints should be considered experimentally. Moreover, perturbations can also be performed within a single subcarrier and phenomenological approaches extracting features from the perturbations and employing regressions or machine learning. 

Additionally, perturbations can also be engineered to enhance the cyclo-stationary spectral region of the signal similarly to \cite{Xu2015}, this can be performed while maintaining the match-filtering properties. This is congruent with the frequency domain description of perturbations covered in this work and it can be considered as future work.

\section*{Acknowledgments}
The authors would like to gratefully acknowledge discussions, ideas and guidance from D. Charlton, C. Laperle, M.E Mousa-Pasandi, M. Hubbard, M. Reimer, and M. O’Sullivan; and thank Ciena for the donation of equipment, funding, as well as technical and measurements support.
\section*{Appendix}
\subsection{Constant Power NLN terms}
Considering the constant power constraint of Equation \ref{eq:ConstPowerContr}, the NSR NLN categories are: 
\begin{equation}
\begin{split}
NSR_{NLNref,2}^{(A)}= NSR^{(A)}_{NLNref}(A,A,A) - 
\\
\frac{K_A}{K_B} NSR^{(A)}_{NLNref}(B,A,A) + \frac{K_A^2}{K_B^2} NSR^{(A)}_{NLNref}(B,B,A) - 
\\
\frac{K_A^3}{K_B^3} NSR^{(A)}_{NLNref}(B,B,B),
\end{split}
\end{equation}
\begin{equation}
\begin{split}
NSR_{NLNref,1}^{(A)}= \frac{1}{K_B} NSR^{(A)}_{NLNref}(B,A,A) + 
\\
\frac{3 K_A^2}{K_B^3} NSR^{(A)}_{NLNref}(B,B,B) -  \frac{K_A}{K_B^2} NSR^{(A)}_{NLNref}(B,B,A),
\end{split}
\end{equation}
\begin{equation}
NSR_{NLNref,0}^{(A)}= \frac{1}{K_B^3}NSR^{(A)}_{NLNref}(B,B,B)
\end{equation}

the APSD NLN categories for a constant launch power considering $F_A$ and $F_B$, in $F_N$ are:
\begin{equation}
\begin{split}
\overline{S}_{NLNref,3}^{(N)} = \overline{S}_{NLNref}^{(N)}(A,A,A)+\frac{K_A^2}{K_B^2} \overline{S}_{NLNref}^{(N)}(B,B,A)  \\
-\frac{K_A^3}{K_B^3} \overline{S}_{NLNref}^{(N)}(B,B,B) - \frac{K_A}{K_B} \overline{S}_{NLNref}^{(N)}(B,A,A),
\end{split}
\end{equation}
\begin{equation}
\begin{split}
\overline{S}_{NLNref,2}^{(N)} = \frac{1}{K_B} \overline{S}_{NLNref}^{(N)}(B,A,A) + 3\frac{K_A^2}{K_B^3} \overline{S}_{NLNref}^{(N)}(B,B,B)\\
\frac{2K_A}{K_B^2} \overline{S}_{NLNref}^{(N)}(B,B,A),
\end{split}
\end{equation}
\begin{equation}
\begin{split}
\overline{S}_{NLNref,1}^{(N)} = \frac{1}{K_B^2} \overline{S}_{NLNref}^{(N)}(B,B,A) - 3\frac{K_A}{K_B^3} \overline{S}_{NLNref}^{(N)}(B,B,B),
\end{split}
\end{equation}
\begin{equation}
\begin{split}
\overline{S}_{NLNref,0}^{(N)} = \frac{1}{K_B^3} \overline{S}_{NLNref}^{(N)}(B,B,B),
\end{split}
\end{equation}

\subsection{Categories Symmetries}
Given the perturbation topology of Figure \ref{fig:PertSimplExampleTotal} and the definition of $|TX_{ref}|^2$, the following symmetries apply to the $F_{BOI}$ generated NLN categories.
\begin{equation}
\begin{split}
NSR^{(A)}_{NLNref}(A,A,A) =NSR^{(B)}_{NLNref}(B,B,B), \\
NSR^{(A)}_{NLNref}(B,A,A) =NSR^{(B)}_{NLNref}(B,B,A), \\
NSR^{(A)}_{NLNref}(B,B,A) =NSR^{(B)}_{NLNref}(B,A,A), \\
NSR^{(A)}_{NLNref}(B,B,B) =NSR^{(B)}_{NLNref}(A,A,A), \\
\end{split}
\end{equation}

For the case of APSDs focusing on $F_N$, the following symmetries exists:
\begin{equation}
\begin{split}
\overline{S}_{NLN_{ref}}^{(N)}(A,A,A)	=	\overline{S}_{NLN_{ref}}^{(N)}(B,B,B),
\\
\overline{S}_{NLN_{ref}}^{(N)}(B,A,A) 	=	\overline{S}_{NLN_{ref}}^{(N)}(B,B,A),
\end{split}
\label{eq:IntraNLNsymFragments}
\end{equation}

For inter-channel categories, the symmetries are:
\begin{equation}
\begin{split}
\overline{S}_{NLN_{ref}}^{(N)}(OB,A,A)	=	\overline{S}_{NLN_{ref}}^{(N)}(OB,B,B),
\\
\overline{S}_{NLN_{ref}}^{(N)}(OB,OB,A)	=	\overline{S}_{NLN_{ref}}^{(N)}(OB,OB,B),
\end{split}
\label{eq:InterNLNsymFragments}	
\end{equation}

\ifCLASSOPTIONcaptionsoff
  \newpage
\fi

\bibliographystyle{ieeetr} 
\bibliography{library}

\end{document}